\documentclass[aps,pra,reprint,groupedaddress,amsmath,amssymb,longbibliography]{revtex4-2}

\pdfpagewidth=\paperwidth
\pdfpageheight=\paperheight

\usepackage{graphicx}

\begin{document}

%\preprint{}

\title{Concatenated Composite Pulses Beyond Local Residual-Error Preservation}


\author{Masamitsu Bando}
\email[Contact author: ]{bando@ktc.ac.jp}
%\thanks{}
\affiliation{Kindai University Technical College, Nabari, Mie 518-0459, Japan}


\date{September 19, 2026}

\begin{abstract}
Residual-error preservation (REP) provides a sufficient rule for constructing
concatenated composite pulses that compensate multiple systematic errors.
We show that local REP is not necessary: deviations from REP can cancel
across the full sequence.
Using first-order error generators, we derive a necessary and sufficient
condition for concatenation to retain the robustness of an outer sequence.
A simple sufficient condition is that all inner pulses rescale the
corresponding elementary error generators by a common real factor,
not necessarily unity.
We illustrate this principle using pulse-length and off-resonance errors,
two systematic error models commonly considered in magnetic resonance.
Short CORPSE and SCROFULOUS each compensate one error without preserving
the other locally.
Combining these inner pulses with suitable equal-rotation-angle outer
sequences yields simultaneous first-order compensation of both errors,
extending concatenated composite-pulse design beyond local REP.
\end{abstract}

%\keywords{}

\maketitle

\section{Introduction}
\label{sec:introduction}

Systematic control errors limit the accuracy of quantum operations even
when the same control sequence is applied reproducibly.
Composite pulses suppress such errors by replacing an elementary operation
with a sequence whose constituent errors cancel while the desired ideal
operation is retained~\cite{Levitt1986,Cummins2003}.
Developed in NMR spectroscopy, these techniques also provide tools for
accurate quantum logic~\cite{Cummins2000,Jones2011}.
In magnetic resonance, two common models are pulse-length errors (PLEs),
which describe fractional errors in the rotation angle, and off-resonance
errors (OREs), which arise from detuning between the drive and the
resonance frequency.
Representative families include BB1~\cite{Wimperis1994} and
SK1~\cite{Brown2004} for PLE compensation, and
CORPSE~\cite{Cummins2003} for ORE compensation.
Systematic constructions also extend error cancellation to arbitrarily
high orders~\cite{Brown2004,Alway2007,Low2014}.
A sequence that compensates one error type, however, need not compensate
another. Combining different compensation mechanisms therefore requires
control of the residual error as well as the ideal operation.

Concatenated composite pulses (CCCPs) address this problem by replacing
the elementary pulses of an outer composite sequence with inner composite
pulses that compensate a different error.
The first-order error-composition framework of
Ref.~\cite{Ichikawa2011} provided a construction with CORPSE inner pulses
and a SCROFULOUS outer sequence.
Residual-error preservation (REP) subsequently supplied a general
prescription for such combinations~\cite{Bando2013}.
In the REP-based construction, each inner pulse implements the ideal
operation of the pulse it replaces and preserves its first-order response
to the error compensated by the outer sequence.
The outer cancellation then survives the replacement, while the inner
pulses compensate the other error.
The utility of CCCPs has also been demonstrated in liquid-state NMR
spectroscopy~\cite{Bando2020}.

REP imposes this matching locally, on each replacement.
Cancellation over the full sequence need not require every local
response to be preserved.
Short CORPSE and SCROFULOUS, for example, compensate OREs and PLEs,
respectively, but lack REP for the remaining error
type~\cite{Cummins2003,Bando2013}.
Under what conditions can such non-REP inner pulses nevertheless be
combined with an outer sequence to compensate both errors?
Cancellation of residual errors across a sequence is already an
established design principle. Direct designs yield composite refocusing
pulses and short NOT gates that compensate both
errors~\cite{Odedra2012,Jones2013}, while planar geometric constructions
can incorporate CORPSE nesting for arbitrary single-qubit
operations~\cite{Ichikawa2014}.
Recent work has also developed composite quantum gates with simultaneous
compensation of multiple control errors using derivative cancellation
and numerical optimization~\cite{Tonchev2026}.
Nested antisymmetric NOT gates also use an outer phase pattern to
cancel residual errors of inner blocks~\cite{JonesNested2013}.
Our question concerns the compatibility of inner and outer operations
when the local REP premise of concatenation is relaxed.

We answer this question using right error generators and the established
first-order composition rule of Ref.~\cite{Ichikawa2011}.
Comparing the residual responses before and after replacement gives a
necessary and sufficient condition for retaining the first-order
robustness of an outer sequence: deviations from local REP must sum to
zero after conjugation by the preceding ideal operations.
We call this the global REP condition.
A simple sufficient condition is that every inner residual generator
equals the corresponding elementary generator multiplied by a common
real factor, which need not be unity.
Short CORPSE and SCROFULOUS realize this structure with the roles of
PLE and ORE interchanged. For each family, suitable equal-rotation-angle
outer sequences make the factor common and yield simultaneous
first-order compensation.
The contribution is thus a compatibility criterion and explicit
constructions beyond local REP, rather than a new error-composition rule.

Our analysis concerns shared systematic error parameters and
first-order cancellation. The general criterion does not depend on a
particular control Hamiltonian; the explicit single-qubit constructions
use a control and error model commonly considered in NMR.
Time-dependent noise and decoherence require additional analysis:
noise correlations can affect composite-pulse
performance~\cite{Mottonen2006}, while dynamically corrected
gates~\cite{Khodjasteh2009}, robust dynamical
decoupling~\cite{Ryan2010}, and protected-gate
implementations~\cite{Souza2012} address related control problems.
Section~\ref{sec:error-model} defines the error generators and local REP.
Section~\ref{sec:concatenation} derives the global condition and the
common-factor criterion, and Sec.~\ref{sec:constructions} develops
the short CORPSE and SCROFULOUS constructions.
Section~\ref{sec:performance} examines their finite-error fidelity and
implementation cost. Section~\ref{sec:conclusions} discusses the
scope of the results and further design possibilities.

\section{Error Generators and Local Residual-Error Preservation}
\label{sec:error-model}

\subsection{Unitary operations and systematic errors}
\label{subsec:right-generators}

Let $U(x,y)$ be a unitary operation on a fixed finite-dimensional
Hilbert space, with ideal value $U_0=U(0,0)$ and identity operator $I$.
The real parameters $x$ and $y$ describe two systematic control errors.
All intended control settings are held fixed, and the same error
parameters apply to every constituent operation in a sequence.
We assume twice continuous differentiability near $(x,y)=(0,0)$.
No rotation axis, pulse Hamiltonian, or physical realization is specified
at this stage. Section~\ref{sec:constructions} specializes this framework
to single-qubit rotations with pulse-length and off-resonance errors.
Products act on states from right to left, so the rightmost factor
is applied first.

With $y=0$, write $U(x)=U(x,0)$. The right error generator is
\begin{equation}
  K_x[U]=iU_0^\dagger
    \left.\frac{\partial U(x)}{\partial x}\right|_{x=0},
  \label{eq:right-generator}
\end{equation}
where $\dagger$ denotes Hermitian conjugation.
The argument $U$ specifies the full error-dependent family, not just
its ideal value. Distinct implementations of the same ideal operation
can therefore have different generators.
Taylor expansion gives
\begin{equation}
  U(x)=U_0\bigl(I-ixK_x[U]\bigr)+O(x^2).
  \label{eq:one-error-expansion}
\end{equation}
The remainder is understood in a matrix norm. Unitarity implies that
$U_0^\dagger\partial_xU|_0$ is anti-Hermitian, so $K_x[U]$ is Hermitian.

Equation~\eqref{eq:one-error-expansion} describes the relative error
$U_0^\dagger U(x)$, not the error term in the physical Hamiltonian.
No commutativity between $U_0$ and $K_x[U]$ is assumed.
If the error factor is instead placed on the left, its generator is
\begin{align}
  L_x[U]&=U_0K_x[U]U_0^\dagger,\notag\\
  U(x)&=\bigl(I-ixL_x[U]\bigr)U_0+O(x^2).
  \label{eq:left-generator}
\end{align}
We use the right-factor convention throughout.

Defining $K_y[U]$ analogously, with both derivatives evaluated at the
error-free point, gives
\begin{align}
  U(x,y)
  &=U_0\bigl(I-ixK_x[U]-iyK_y[U]\bigr)\notag\\
  &\quad+O(x^2,xy,y^2).
  \label{eq:two-error-expansion}
\end{align}
The remainder has norm bounded by a constant times
$x^2+|xy|+y^2$ near the origin.
We call $U$ first-order robust against $x$ when $K_x[U]=0$.
Simultaneous first-order robustness means
\begin{equation}
  K_x[U]=K_y[U]=0.
  \label{eq:simultaneous-robustness}
\end{equation}
These statements use fixed phase representatives of the unitary
operations. Error-independent global phases leave the generators
unchanged. If robustness is instead required only up to an
error-dependent global phase, identity components of the generators
must be disregarded consistently. For the pulse models used below,
all first-order generators are traceless, so this distinction does
not affect the results.

\subsection{Local residual-error preservation}
\label{subsec:local-rep}

Let $R(x,y)$ be a reference elementary operation and
$U_{\mathrm{CP}}(x,y)$ a composite implementation of the same ideal
operation. The symbol $R$ here denotes a general unitary family;
its explicit rotation form is introduced only in
Sec.~\ref{sec:constructions}.
Their error-free values are denoted by $R_0=R(0,0)$ and
$U_{\mathrm{CP},0}=U_{\mathrm{CP}}(0,0)$.
We align error-independent global phases so that
$U_{\mathrm{CP},0}=R_0$.
The composite implementation has local REP with respect to $x$ if
\begin{equation}
  K_x[U_{\mathrm{CP}}]=K_x[R].
  \label{eq:local-rep}
\end{equation}
With the common ideal reference, this equality is equivalent to
equality of the first-order error terms, as in the original REP
definition~\cite{Bando2013}.
In concatenation, the inner pulse compensates one error, while REP
concerns its response to the other, residual error. REP preserves
that response; it does not require it to vanish.
For the applications below, REP-PLE and REP-ORE denote this property
with respect to pulse-length and off-resonance errors, respectively.

\section{Concatenation Beyond Local REP}
\label{sec:concatenation}

\subsection{Composition of first-order error generators}
\label{subsec:generator-composition}

We use the first-order robustness framework of
Ref.~\cite{Ichikawa2011}. Its Eq.~(29) expresses the total error as a
sum of interaction-picture error contributions conjugated by the
preceding ideal operations, with vanishing total error as the
robustness condition. We reproduce this established composition rule
in right-generator notation, then compare an outer sequence with its
concatenated counterpart to identify conditions beyond local REP.

Consider an outer sequence of $N$ elementary pulses, numbered in their
order of application. Fix a shared error parameter $x$, set all
other errors to zero, and write
\begin{align}
  V(x)&=R_N(x)\cdots R_1(x),\notag\\
  R_{j,0}&=R_j(0).
  \label{eq:outer-sequence}
\end{align}
Here $R_j(x)$ denotes the $j$th reference unitary family in the
sense of Sec.~\ref{sec:error-model}. All control settings are fixed.
Concatenation replaces each $R_j(x)$ by an inner composite pulse $U_j(x)$
implementing the same ideal operation. Write $U_{j,0}=U_j(0)$ and
$V_0=V(0)$. As in Sec.~\ref{subsec:local-rep},
we align any error-independent global phases so that
$U_{j,0}=R_{j,0}$. This alignment does not change the right generators.
The resulting sequence and its ideal value $W_0=W(0)$ are
\begin{align}
  W(x)&=U_N(x)\cdots U_1(x),\notag\\
  W_0&=V_0=R_{N,0}\cdots R_{1,0}.
  \label{eq:inner-sequence}
\end{align}
Define the ideal evolution through the first $j$ pulses by
\begin{align}
  P_0&=I,\notag\\
  P_j&=R_{j,0}\cdots R_{1,0},\qquad j=1,\ldots,N.
  \label{eq:prefix-evolution}
\end{align}
Then $P_{j-1}$ is the ideal evolution preceding the $j$th pulse,
and $P_N=V_0=W_0$. Since $U_{j,0}=R_{j,0}$ for every replacement,
the same $P_j$ describes the ideal evolution through the first $j$
inner blocks of the concatenated sequence.

The product rule gives
\begin{equation}
  \left.\partial_x W\right|_0
  =\sum_{j=1}^N R_{N,0}\cdots R_{j+1,0}
    \left.\partial_x U_j\right|_0 P_{j-1},
  \label{eq:product-derivative}
\end{equation}
where an empty product is the identity.
From Eq.~\eqref{eq:right-generator},
$\left.\partial_x U_j\right|_0=-iR_{j,0}K_x[U_j]$.
Multiplication of Eq.~\eqref{eq:product-derivative} by $iW_0^\dagger$
therefore yields the composition rule
\begin{equation}
  K_x[W]=\sum_{j=1}^N P_{j-1}^\dagger K_x[U_j]P_{j-1}.
  \label{eq:generator-composition}
\end{equation}
For a single error parameter $x$, the quantities $\delta W_I^j$
and $\Delta W$ in Ref.~\cite{Ichikawa2011} correspond at first order
to $xK_x[U_j]$ and $xK_x[W]$, respectively, when each inner block is
treated as one segment. The ideal prefix $V^{j-1}$ of that reference
corresponds to $P_{j-1}$ here.
The same calculation for the elementary outer sequence gives
\begin{equation}
  K_x[V]=\sum_{j=1}^N P_{j-1}^\dagger K_x[R_j]P_{j-1}.
  \label{eq:outer-generator}
\end{equation}
The conjugation by $P_{j-1}$ expresses each local generator in a common
frame referenced to the beginning of the sequence. Thus the total error
is determined by a sum of conjugated generators, not by an untransformed
sum of local errors. No commutation between a generator and an ideal
pulse is required.
For single-qubit sequences of $\pi$ pulses, this frame transformation can also be
expressed through toggling-frame phases~\cite{Jones2013,JonesNested2013}.
Here the operator form allows us to compare elementary and composite
implementations without restricting every target rotation to $\pi$.

\subsection{Global cancellation condition}
\label{subsec:global-cancellation}

The following criterion is a direct consequence of the composition
rule. Its role is to distinguish preservation of the full sequence's
response from preservation in each individual replacement.
For each replacement, define the difference between the inner and
elementary generators by
\begin{equation}
  \Delta K_{x,j}=K_x[U_j]-K_x[R_j].
  \label{eq:generator-difference}
\end{equation}
Subtracting Eq.~\eqref{eq:outer-generator} from
Eq.~\eqref{eq:generator-composition} gives
\begin{equation}
  K_x[W]=K_x[V]
    +\sum_{j=1}^N P_{j-1}^\dagger\Delta K_{x,j}P_{j-1}.
  \label{eq:concatenation-difference}
\end{equation}
Consequently, if the outer sequence is first-order robust against $x$,
then the concatenated sequence is first-order robust against $x$ if and
only if
\begin{equation}
  \sum_{j=1}^N P_{j-1}^\dagger\Delta K_{x,j}P_{j-1}=0.
  \label{eq:global-rep}
\end{equation}
We call Eq.~\eqref{eq:global-rep} the \emph{global REP condition}.
It preserves the first-order error response of the complete sequence,
rather than requiring its preservation in every constituent replacement.
The assumption $K_x[V]=0$ is essential to its interpretation as a
criterion for robustness: without that assumption,
Eq.~\eqref{eq:global-rep} implies $K_x[W]=K_x[V]$, which need not vanish.

Local REP imposes $\Delta K_{x,j}=0$ for every $j$ and is therefore a
sufficient condition for Eq.~\eqref{eq:global-rep}.
It is not necessary, since nonzero differences can cancel after
conjugation. The condition concerns the matrix sum, including the
directions and signs of its terms; it does not require any individual
difference to be small. The constructions in Sec.~\ref{sec:constructions}
provide explicit realizations with non-REP inner pulses.

To account for the second shared error parameter $y$, suppose each
inner pulse is first-order robust against $y$, so that
$K_y[U_j]=0$ for all $j$. The same composition rule then gives
\begin{equation}
  K_y[W]=\sum_{j=1}^N P_{j-1}^\dagger K_y[U_j]P_{j-1}=0.
  \label{eq:inner-error-cancellation}
\end{equation}
Thus, under inner robustness against $y$ and outer robustness against $x$,
the global REP condition is necessary and sufficient for simultaneous
first-order robustness of the concatenated sequence.
This statement uses the shared systematic error parameters of
Sec.~\ref{sec:error-model}; it does not assert cancellation of
mixed or higher-order terms.

\subsection{A common proportionality factor}
\label{subsec:common-factor}

A useful sufficient condition can be imposed directly on the residual
generators. Fix the error $x$ and the outer sequence, and suppose there
is a real number $q$, common to all replacements, such that
\begin{equation}
  K_x[U_j]=qK_x[R_j]\qquad (j=1,\ldots,N).
  \label{eq:common-factor}
\end{equation}
The factor is evaluated at the error-free point and may depend on the
chosen pulse construction and fixed rotation parameters, but not on $j$.
It need not be positive.
Substitution into Eq.~\eqref{eq:generator-composition} gives
\begin{align}
  K_x[W]
  &=q\sum_{j=1}^N P_{j-1}^\dagger K_x[R_j]P_{j-1}\notag\\
  &=qK_x[V].
  \label{eq:scaled-outer-generator}
\end{align}
Hence any first-order robustness of the outer sequence against $x$ is
retained after these replacements. Equivalently,
$\Delta K_{x,j}=(q-1)K_x[R_j]$ implies
\begin{equation}
  \sum_{j=1}^N P_{j-1}^\dagger\Delta K_{x,j}P_{j-1}
  =(q-1)K_x[V]=0
  \label{eq:common-factor-global-rep}
\end{equation}
when $K_x[V]=0$. If the inner pulses also compensate the other error,
Eq.~\eqref{eq:inner-error-cancellation} completes the proof of
simultaneous first-order robustness.

The choice $q=1$ recovers local REP. For $q\ne1$, a replacement with
$K_x[R_j]\ne0$ violates local REP, yet remains compatible with global
cancellation. Negative $q$ includes reversal of the residual generator,
not merely a change in its magnitude. If $K_x[R_j]=0$, the proportionality
condition requires $K_x[U_j]=0$ and that element alone does not determine
$q$.

For a rotation-based inner-pulse family whose residual generator is
proportional to the elementary one with a factor depending only on its
rotation angle, equal outer rotation angles make this factor common.
This conclusion presumes the same branch and error model for all inner
pulses. Section~\ref{sec:constructions} establishes this structure for
short CORPSE with residual PLE and for SCROFULOUS with residual ORE.
Equal rotation angles alone do not guarantee the condition for an
arbitrary inner-pulse family.

Neither a common proportionality factor nor equal rotation angles are
necessary for Eq.~\eqref{eq:global-rep}. Different factors, or even local
generators that are not proportional to their elementary counterparts,
are admissible whenever the conjugated differences sum to zero.
The common-factor condition is therefore a construction principle within
the general cancellation criterion, rather than a replacement for it.

\section{Single-Qubit Constructions with Non-REP Inner Pulses}
\label{sec:constructions}

The preceding conditions apply independently of the specific form of the
reference operations. We now choose a single-qubit control model and
construct two complementary examples with non-REP inner pulses.

\subsection{Elementary rotations and systematic errors}
\label{subsec:elementary-errors}

We now apply the general framework to single-qubit control.
We first define ideal rotations about arbitrary axes, then specialize
to a control and error model commonly used in NMR.
Let $I$ denote the $2\times2$ identity matrix and
$\sigma_x,\sigma_y,\sigma_z$ the Pauli matrices, with
$\sigma_x\sigma_y=i\sigma_z$ and cyclic permutations.
Writing $\boldsymbol{\sigma}=(\sigma_x,\sigma_y,\sigma_z)$, a rotation
about an arbitrary unit vector $\boldsymbol n=(n_x,n_y,n_z)$ is
\begin{align}
  R_{\boldsymbol n}(\theta)
  &=\exp\left[-i\frac{\theta}{2}\boldsymbol n\cdot\boldsymbol{\sigma}\right]
    \notag\\
  &=\cos\frac{\theta}{2}I
    -i\sin\frac{\theta}{2}\boldsymbol n\cdot\boldsymbol{\sigma},
  \qquad |\boldsymbol n|=1.
  \label{eq:general-rotation}
\end{align}
For the explicit constructions, we adopt a control model commonly
used in NMR~\cite{Claridge1999}: ideal pulses generate rotations about axes in the
transverse plane, while pulse-length and off-resonance errors describe
systematic deviations from the intended control~\cite{Cummins2003,Ichikawa2011}.
This specialization is not required by the general theory developed
above. The constructions also apply to other physical platforms when
their effective control and error models have the same form.

The following pulse constructions use axes in the $xy$ plane,
$\boldsymbol n_\phi=(\cos\phi,\sin\phi,0)$.
For an azimuthal angle $\phi$, define
\begin{equation}
  \sigma_\phi=\cos\phi\,\sigma_x+\sin\phi\,\sigma_y.
  \label{eq:rotation-axis}
\end{equation}
We abbreviate $R_{\boldsymbol n_\phi}(\theta)$ as $R(\theta,\phi)$.
An ideal elementary pulse in this control model is
\begin{align}
  R(\theta,\phi)
  &=\exp\left(-i\frac{\theta}{2}\sigma_\phi\right)\notag\\
  &=\cos\frac{\theta}{2}\,I-i\sin\frac{\theta}{2}\,\sigma_\phi.
  \label{eq:ideal-rotation}
\end{align}
The restriction to transverse axes is a choice of available elementary
controls, not an assumption in Secs.~\ref{sec:error-model}
and~\ref{sec:concatenation}. The $n_z$ component in
Eq.~\eqref{eq:general-rotation} describes an ideal rotation axis;
it is distinct from the detuning error introduced below.

The real parameters $\theta$ and $\phi$ are measured in radians;
$\theta$ is the rotation angle in the absence of errors, and $\phi$ is
defined modulo $2\pi$. Pulse products act on states from right to left,
so the rightmost factor is applied first.
Equality of ideal gates up to a global phase does not imply equality of
their error responses. In particular, we do not identify different rotation
angles merely because they implement the same ideal gate.

We denote an imperfect pulse by $R(\theta,\phi;\varepsilon,f)$.
The ideal operation $R(\theta,\phi)$ is recovered at $\varepsilon=f=0$.
The dimensionless real parameters $\varepsilon$ and $f$ describe PLE
and ORE, respectively. These are the specific choices of $x$ and $y$
used to illustrate the general theory. With the other error set to zero, the models
are~\cite{Cummins2003,Bando2013}
\begin{align}
  R(\theta,\phi;\varepsilon,0)
  &=R((1+\varepsilon)\theta,\phi),
  \label{eq:ple-model}\\
  R(\theta,\phi;0,f)
  &=\exp\left[-i\frac{\theta}{2}
    (\sigma_\phi+f\sigma_z)\right].
  \label{eq:ore-model}
\end{align}
Here $f=\Delta/\Omega$, where $\Delta$ is the signed detuning and
$\Omega>0$ is the error-free Rabi angular frequency.
The sign convention for $\Delta$ is fixed by the coefficient of $\sigma_z$
in Eq.~\eqref{eq:ore-model}.
We hold $\theta$, $\phi$, and $\Omega$ fixed when varying the error parameters.
The same $\varepsilon$ and $f$ apply to every constituent pulse;
the constructions considered below also use a common $\Omega$.

For the first-order analysis, the two-parameter notation denotes a unitary
family that is twice continuously differentiable near $(0,0)$ and obeys
Eqs.~\eqref{eq:ple-model} and~\eqref{eq:ore-model} on the coordinate axes.
These restrictions determine its first derivatives, but not its mixed
error terms. No particular extension away from the axes is needed for
the first-order results below. For finite-error evaluation, we specify
the simultaneous-error model in Sec.~\ref{subsec:fidelity}.

\subsection{Elementary error generators}
\label{subsec:elementary-generators}

We evaluate the elementary generators needed to apply the general
REP and common-factor conditions to this control model.
In $K_x[R]$, the argument denotes $R(\theta,\phi;\varepsilon,f)$
at fixed $\theta$ and $\phi$.
Since a PLE changes only the rotation angle about the same axis,
Eq.~\eqref{eq:ple-model} yields
\begin{equation}
  K_\varepsilon[R]=\frac{\theta}{2}\sigma_\phi.
  \label{eq:elementary-ple-generator}
\end{equation}
For ORE, using $(\sigma_\phi+f\sigma_z)^2=(1+f^2)I$ gives
\begin{align}
  R(\theta,\phi;0,f)
  &=\cos\left(\frac{\theta}{2}\sqrt{1+f^2}\right)I\notag\\
  &\quad-i\frac{\sin\left(\frac{\theta}{2}\sqrt{1+f^2}\right)}
                   {\sqrt{1+f^2}}
             (\sigma_\phi+f\sigma_z).
  \label{eq:ore-exact}
\end{align}
Although the scalar trigonometric factors have no linear term in $f$,
the axis operator does. Consequently,
\begin{equation}
  R(\theta,\phi;0,f)
  =R(\theta,\phi)-if\sin\frac{\theta}{2}\,\sigma_z+O(f^2).
  \label{eq:ore-linear}
\end{equation}
Multiplying its derivative by $iR(\theta,\phi)^\dagger$ then gives
\begin{align}
  K_f[R]
  &=\sin\frac{\theta}{2}\cos\frac{\theta}{2}\,\sigma_z
  +\sin^2\frac{\theta}{2}\,\sigma_{\phi+\pi/2},
  \label{eq:elementary-ore-generator}
\end{align}
where $\sigma_{\phi+\pi/2}=\cos\phi\,\sigma_y-\sin\phi\,\sigma_x$.
Unlike the PLE generator, this generator does not in general commute
with the ideal rotation.

\subsection{Short CORPSE inner pulses}
\label{subsec:short-corpse}

For a target $R(\theta,\phi)$ with $0<\theta\leq2\pi$, define
\begin{equation}
  k=\arcsin\left(\frac{\sin(\theta/2)}{2}\right),\quad
  a=\frac{\theta}{2}-k,\quad b=2\pi-2k,
  \label{eq:short-parameters}
\end{equation}
using the principal value of $\arcsin$.
The short CORPSE sequence has pulse angles $(a,b,a)$ and phases
$(\phi,\phi+\pi,\phi)$~\cite{Cummins2003,Bando2013}.
To specify its error-dependent family, let
\begin{align}
  A(\varepsilon,f)&=R(a,\phi;\varepsilon,f),\notag\\
  B(\varepsilon,f)&=R(b,\phi+\pi;\varepsilon,f),\notag\\
  U_{\mathrm{short}}(\varepsilon,f)&=-A(\varepsilon,f)
    B(\varepsilon,f)A(\varepsilon,f).
  \label{eq:short-family}
\end{align}
The constant minus sign aligns the ideal operation with the target:
$A_0B_0A_0=R(\theta-2\pi,\phi)=-R(\theta,\phi)$.
It has no effect on the right error generators.

For PLE alone, all three rotations share an axis up to its sign. Thus
\begin{equation}
  U_{\mathrm{short}}(\varepsilon,0)
  =R(\theta,\phi)
    R\bigl(\varepsilon(\theta-2\pi),\phi\bigr),
  \label{eq:short-ple-exact}
\end{equation}
and the residual PLE generator is
\begin{align}
  K_\varepsilon[U_{\mathrm{short}}]
  &=\frac{\theta-2\pi}{2}\sigma_\phi\notag\\
  &=\left(1-\frac{2\pi}{\theta}\right)K_\varepsilon[R].
  \label{eq:short-residual}
\end{align}
For completeness, its ORE cancellation follows directly from
Eq.~\eqref{eq:ore-linear}. Differentiating $ABA$ at $f=0$ gives
\begin{equation}
  \left.\partial_f(ABA)\right|_0
  =-i\left[2\sin\frac a2\cos\frac{a-b}{2}
        +\sin\frac b2\right]\sigma_z.
  \label{eq:short-ore-derivative}
\end{equation}
The scalar coefficient is $2\sin k-\sin(\theta/2)=0$, so
$K_f[U_{\mathrm{short}}]=0$.

The difference from the elementary PLE generator is
\begin{equation}
  K_\varepsilon[U_{\mathrm{short}}]-K_\varepsilon[R]
  =-\pi\sigma_\phi.
  \label{eq:short-difference}
\end{equation}
It is nonzero throughout the stated domain, including $\theta=2\pi$,
where the residual generator itself vanishes.
For a PLE-robust outer sequence whose elements are replaced by these
inner pulses, Eq.~\eqref{eq:global-rep} therefore reduces to
\begin{equation}
  \sum_{j=1}^N P_{j-1}^\dagger\sigma_{\phi_j}P_{j-1}=0.
  \label{eq:short-global-condition}
\end{equation}
By contrast, outer PLE robustness only guarantees that the same sum
weighted by $\theta_j/2$ vanishes. For unequal angles these are distinct
conditions. If $\theta_j=\alpha$ for every $j$, with
$0<\alpha\leq2\pi$, they coincide up to a constant factor. Equivalently,
Eq.~\eqref{eq:common-factor} holds with $q=1-2\pi/\alpha$.
The concatenated sequence is then first-order robust against both errors.

An equal-angle outer sequence can be obtained by splitting a known
PLE-robust sequence whenever each constituent angle is a positive
integer multiple of a common angle $\alpha$.
For $\theta_j=m_j\alpha$, $m_j\in\mathbb{N}$, the PLE model obeys
\begin{equation}
  R(\theta_j,\phi_j;\varepsilon,0)
  =\bigl[R(\alpha,\phi_j;\varepsilon,0)\bigr]^{m_j}.
  \label{eq:ple-splitting}
\end{equation}
The split pieces are consecutive pulses of the same phase and Rabi
frequency, without intervening delays, and share the same systematic
error. This replacement preserves the complete PLE-dependent outer
operation, not just its first-order term. After splitting, the pieces
are renumbered in time order, and $N$ and $P_{j-1}$ refer to this new
outer sequence. Replacing every piece by short CORPSE then gives a
concatenation with short CORPSE inner pulses and a split outer sequence.

Two examples use SK1~\cite{Brown2004,Bando2013} and
BB1~\cite{Wimperis1994,Bando2013}.
Set $\chi=\arccos[-\theta/(4\pi)]$, using the principal value, for
$0<\theta\leq2\pi$. In time order, the SK1 parameters are
\begin{equation}
  (\theta,\phi),\quad(2\pi,\phi-\chi),\quad(2\pi,\phi+\chi).
  \label{eq:sk1-parameters}
\end{equation}
For BB1, we use the form with the correction pulses preceding the
target pulse:
\begin{equation}
  (\pi,\phi+\chi),\quad(2\pi,\phi+3\chi),\quad
  (\pi,\phi+\chi),\quad(\theta,\phi).
  \label{eq:bb1-parameters}
\end{equation}
Both ideal products implement $R(\theta,\phi)$ and cancel first-order
PLE. At $\theta=\pi$, splitting each $2\pi$ pulse into two $\pi$
pulses produces a five-pulse equal-angle outer sequence in either case.
Here $q=-1$: each short CORPSE block reverses the elementary PLE
generator, yet their conjugated sum still vanishes. Each construction
contains 15 elementary pulses before any merging of adjacent pulses.

For these SK1 and BB1 forms, a finite exact equal-angle splitting exists
when $\theta/\pi$ is rational. This commensurability condition belongs
to the splitting procedure, not to the general cancellation criterion.
Approximating an irrational target angle by a rational one introduces
an additional ideal-gate approximation error. Finer splitting also
increases the number of inner blocks; first-order cancellation alone
does not imply improved performance at fixed nonzero errors.

\subsection{SCROFULOUS inner pulses}
\label{subsec:scrofulous}

SCROFULOUS provides the complementary construction: it compensates
PLE but does not preserve the elementary ORE response locally.
We use the symmetric three-pulse family of
Ref.~\cite{Cummins2003} on the branch $0<\theta\leq\pi$.
Let $a\in[\pi/2,\pi]$ be the unique solution of
\begin{equation}
  \frac{\sin a}{a}=\frac{2\cos(\theta/2)}{\pi},
  \label{eq:scrof-angle}
\end{equation}
and define, with principal values of $\arccos$,
\begin{align}
  \delta&=\arccos\left(-\frac{\pi}{2a}\right),\notag\\
  \beta&=\arccos\left[-\frac{\pi\cos a}
                              {2a\sin(\theta/2)}\right],\notag\\
  \varphi_1&=\phi+\beta,\qquad
  \varphi_2=\phi+\beta-\delta.
  \label{eq:scrof-phases}
\end{align}
Here $a$ is local to the SCROFULOUS definition and is distinct from
the short CORPSE angle in Eq.~\eqref{eq:short-parameters}.
With
\begin{align}
  A(\varepsilon,f)&=R(a,\varphi_1;\varepsilon,f),\notag\\
  B(\varepsilon,f)&=R(\pi,\varphi_2;\varepsilon,f),\notag\\
  U_{\mathrm{SCROF}}(\varepsilon,f)&=A(\varepsilon,f)
      B(\varepsilon,f)A(\varepsilon,f),
  \label{eq:scrof-family}
\end{align}
these parameters give $U_{\mathrm{SCROF},0}=R(\theta,\phi)$.
The PLE generator vanishes because the composition rule gives
\begin{equation}
  A_0K_\varepsilon[U_{\mathrm{SCROF}}]A_0^\dagger
  =\left(a\cos\delta+\frac\pi2\right)\sigma_{\varphi_2}=0.
  \label{eq:scrof-ple-cancellation}
\end{equation}
This follows from the reflection of $\sigma_{\varphi_1}$ by the
central $\pi$ rotation and $\cos\delta=-\pi/(2a)$.

To derive the residual ORE generator, set $\varepsilon=0$ and write
$s=\sin(a/2)$. Equation~\eqref{eq:ore-linear} gives
\begin{equation}
  \left.\partial_f A\right|_0=-is\sigma_z,\qquad
  \left.\partial_f B\right|_0=-i\sigma_z.
  \label{eq:scrof-local-derivatives}
\end{equation}
Anticommutation of $\sigma_z$ with each transverse axis implies
\begin{align}
  A_0\sigma_z A_0&=\sigma_z,\notag\\
  \sigma_zB_0A_0+A_0B_0\sigma_z
    &=-2s\cos\delta\,\sigma_z.
  \label{eq:scrof-identities}
\end{align}
Differentiating the product in Eq.~\eqref{eq:scrof-family} therefore
yields
\begin{align}
  \left.\partial_f U_{\mathrm{SCROF}}\right|_0
  &=-ig(\theta)\sigma_z,\notag\\
  g(\theta)&=1+\frac\pi a\sin^2\frac a2.
  \label{eq:scrof-ore-derivative}
\end{align}
Comparing with the elementary ORE generator gives
\begin{align}
  K_f[U_{\mathrm{SCROF}}]
  &=g(\theta)R(\theta,\phi)^\dagger\sigma_z\notag\\
  &=q(\theta)K_f[R],\notag\\
  q(\theta)&=\frac{1+(\pi/a)\sin^2(a/2)}{\sin(\theta/2)}.
  \label{eq:scrof-residual}
\end{align}
Throughout this branch, $q(\theta)>1$ and $K_f[R]\ne0$, so local
REP-ORE is absent. In particular, $a=\pi$ and $q=2$ at $\theta=\pi$.

Consider any ORE-robust outer sequence with equal rotation angles
$\alpha$, where $0<\alpha\leq\pi$.
Replacing each pulse by the SCROFULOUS family above, on the same branch,
gives $K_f[U_j]=q(\alpha)K_f[R_j]$ for every $j$.
Section~\ref{subsec:common-factor} then gives
$K_f[W]=q(\alpha)K_f[V]=0$, while
$K_\varepsilon[W]=0$ follows from the PLE robustness of each inner block.

An equal-angle outer sequence can again be obtained by splitting a
known robust sequence, now chosen to compensate ORE. For
$\theta_j=m_j\alpha$, the ORE model obeys the exact identity
\begin{equation}
  R(\theta_j,\phi_j;0,f)
  =\bigl[R(\alpha,\phi_j;0,f)\bigr]^{m_j}.
  \label{eq:ore-splitting}
\end{equation}
Consecutive pieces share the same phase, Rabi frequency, and error
parameters, without intervening delays. Splitting therefore preserves
the entire ORE-dependent outer operation. Replacing every piece by
SCROFULOUS on the same branch gives simultaneous first-order
compensation, just as splitting a PLE-robust outer sequence permits
short CORPSE inner pulses in Sec.~\ref{subsec:short-corpse}.

This construction differs from the direct reversal considered in
Ref.~\cite{Ichikawa2011}, where three SCROFULOUS blocks forming an
unsplit CORPSE outer sequence were found not to compensate both errors.
Here the outer sequence is first split into equal-angle pieces, and
each piece is then replaced by SCROFULOUS. The splitting leaves the
outer operation unchanged but makes the residual proportionality
factor common after replacement. Thus the present construction does
not contradict the failure of that unsplit concatenation.

CORPSE and short CORPSE provide ORE-robust outer sequences for this
construction~\cite{Cummins2003,Bando2013}. To make the angle condition
explicit, set $u=\theta/2-k$ and $v=2\pi-2k$, with $k$ defined in
Eq.~\eqref{eq:short-parameters}. Their pulse angles in time order can
be chosen as $(2\pi+u,v,u)$ and $(u,v,u)$, respectively, with phases
$(\phi,\phi+\pi,\phi)$. For either choice, an exact finite equal-angle
splitting requires that every pulse angle be a positive integer
multiple of a common $\alpha$. A further common subdivision can
ensure $\alpha\leq\pi$, as required by the inner SCROFULOUS branch.
Unlike the BB1 and SK1 examples, these angles also involve
$\arcsin[\sin(\theta/2)/2]$. Rationality of $\theta/\pi$ alone is
therefore not sufficient to establish such a splitting. The design
principle applies whenever the constituent angles are commensurate;
it is not an assertion that every target angle admits a finite exact
split of these particular outer sequences.

For a target $R(\pi,\phi)$, short CORPSE has angles
$(\pi/3,5\pi/3,\pi/3)$. Choosing $\alpha=\pi/3$ gives seven
equal-angle outer pulses, with phases in time order
\begin{equation}
  (\phi,\underbrace{\phi+\pi,\ldots,\phi+\pi}_{5\ \mathrm{pulses}},\phi).
  \label{eq:split-short-outer}
\end{equation}
The split outer sequence retains the ideal target up to a global
phase and remains ORE robust. Substituting SCROFULOUS for each pulse
gives a 21-pulse concatenation before merging, with common residual
factor $q(\pi/3)>1$. Alternatively, the CORPSE angles
$(7\pi/3,5\pi/3,\pi/3)$ split into 13 pulses of angle $\pi/3$,
with phase groups of lengths 7, 5, and 1. Their SCROFULOUS
replacement gives a 39-pulse concatenation before merging.

Both examples implement the same construction principle: equal-angle
splitting makes a non-REP inner response compatible with the outer
cancellation. Neither preservation of the magnitude nor preservation
of the sign of an individual residual generator is required:
SCROFULOUS has $q(\alpha)>1$, whereas the split short CORPSE inner
examples at $\alpha=\pi$ have $q=-1$.
These constructions realize the common-factor condition of
Sec.~\ref{subsec:common-factor} with non-REP inner pulses.
We next examine representative finite-error responses and the
implementation costs of the explicit sequences.

\begin{figure*}[t]
  \centering
  \begin{tabular}{@{}c@{\hspace{3pt}}c@{\hspace{3pt}}c@{}}
    \includegraphics[scale=0.29]{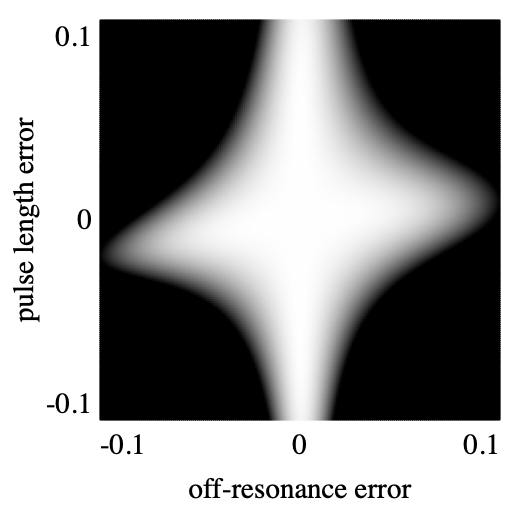} &
    \includegraphics[scale=0.29]{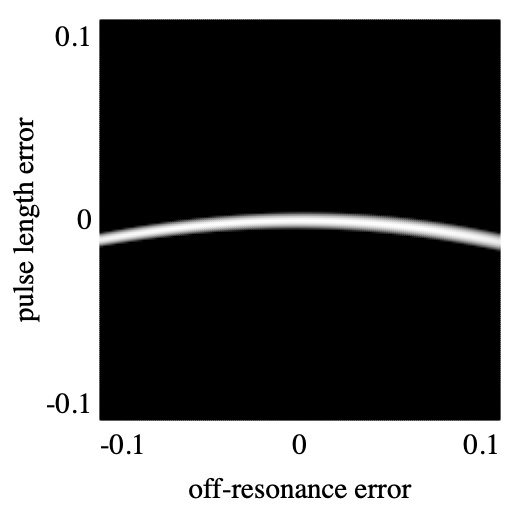} &
    \includegraphics[scale=0.29]{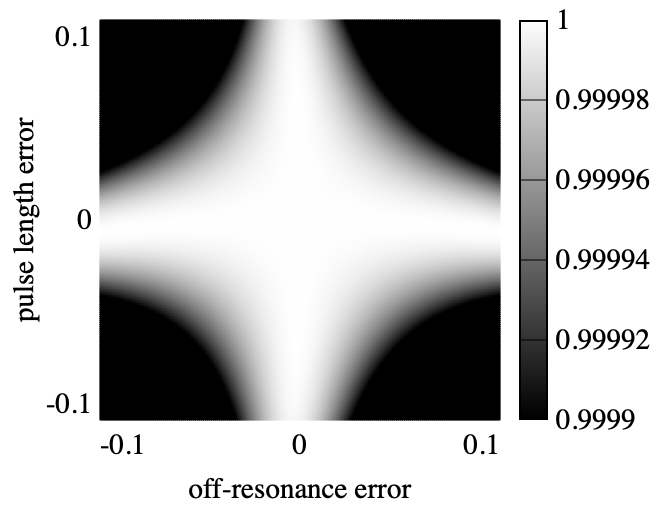} \\
    \small (a) CORPSE/BB1 &
    \small (b) shortCORPSE/BB1 &
    \small (c) shortCORPSE/splitBB1 \\[8pt]
    \includegraphics[scale=0.29]{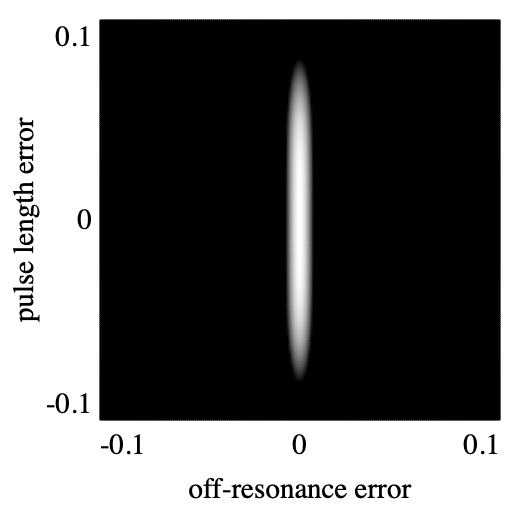} &
    \includegraphics[scale=0.29]{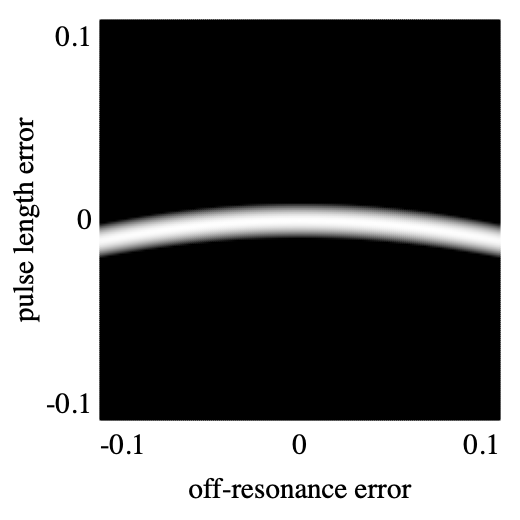} &
    \includegraphics[scale=0.29]{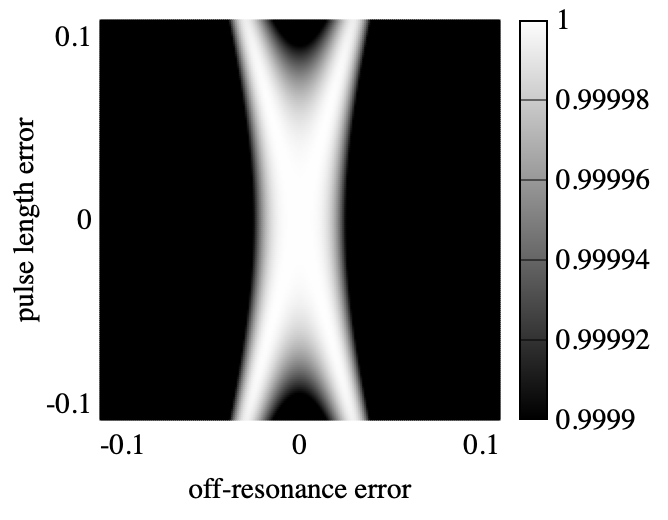} \\
    \small (d) SCROFULOUS &
    \small (e) splitShortCORPSE &
    \small (f) SCROFULOUS/splitShortCORPSE
  \end{tabular}
  \caption{Trace fidelity $F$ of Eq.~\eqref{eq:trace-fidelity} for
  the target $R(\pi,0)$ under the simultaneous-error model of
  Eq.~\eqref{eq:simultaneous-model}. The horizontal and vertical
  axes are ORE $f$ and PLE $\varepsilon$, respectively, both ranging
  from $-0.1$ to $0.1$ in steps of $0.001$, giving a
  $201\times201$ grid in every panel.
  Concatenation labels follow the inner/outer
  convention. All six panels share the grayscale shown at the right
  of each row: black denotes $F\leq0.9999$ and white denotes $F=1$.
  In (c), split BB1 consists of five $\pi$ pulses.
  In (e) and (f), short CORPSE is split into seven pulses of angle
  $\pi/3$; in (f), each is replaced by SCROFULOUS, giving 21
  elementary pulses before merging.}
  \label{fig:fidelity-maps}
\end{figure*}

\section{Performance and Implementation Cost}
\label{sec:performance}

\subsection{Fidelity at finite error strengths}
\label{subsec:fidelity}

The first-order conditions established above concern derivatives at the
error origin. To evaluate the full operation at finite errors, we use
\begin{equation}
  R(\theta,\phi;\varepsilon,f)
  =\exp\left[-\frac{i\theta(1+\varepsilon)}{2}
    (\sigma_\phi+f\sigma_z)\right].
  \label{eq:simultaneous-model}
\end{equation}
Here PLE multiplies the entire rotation angle, including the detuning
contribution. With $\theta=\Omega t$ and $f=\Delta/\Omega$, this
model can be realized by replacing the pulse duration $t$ by
$(1+\varepsilon)t$ while keeping $\Omega$ and $\Delta$ fixed.
It reduces to Eqs.~\eqref{eq:ple-model} and~\eqref{eq:ore-model}
on the respective error axes, so all first-order results remain valid.

Define $\rho=\sqrt{1+f^2}$ and
$\eta=\theta(1+\varepsilon)\rho/2$. The exact elementary operation is
\begin{equation}
  R(\theta,\phi;\varepsilon,f)
  =\cos\eta\,I-i\frac{\sin\eta}{\rho}
    (\sigma_\phi+f\sigma_z).
  \label{eq:simultaneous-exact}
\end{equation}
For a sequence of $M$ elementary pulses with angles $\vartheta_\ell$
and phases $\psi_\ell$ listed in time order, the finite-error operation is
\begin{equation}
  W(\varepsilon,f)
  =R(\vartheta_M,\psi_M;\varepsilon,f)\cdots
   R(\vartheta_1,\psi_1;\varepsilon,f).
  \label{eq:finite-sequence}
\end{equation}
The same error parameters are used in every factor. Evaluation of this
product uses Eq.~\eqref{eq:simultaneous-exact} without a Taylor
truncation; it therefore retains higher-order and mixed-error effects.

We use the phase-insensitive trace fidelity
\begin{equation}
  F(\varepsilon,f)=\frac12\left|
    \operatorname{Tr}\bigl[U_{\mathrm{tar}}^\dagger
      W(\varepsilon,f)\bigr]\right|,
  \label{eq:trace-fidelity}
\end{equation}
where $U_{\mathrm{tar}}=R(\theta,\phi)$ is the target operation.
The absolute trace is not squared. For unitary single-qubit operations,
$0\leq F\leq1$, and $F=1$ precisely when the implemented operation
equals the target up to a global phase.

Numerical calculations use version 0.1.13 of the open-source library
\path{@masabando/quantum-gates}, developed by the
author~\cite{BandoQuantumGates2026}. Its sequence evaluator implements
Eq.~\eqref{eq:simultaneous-exact}, and its fidelity routine evaluates
Eq.~\eqref{eq:trace-fidelity}. No local correction to the library's
error model is applied. The physical error parameters remain common
to all elementary pulses.

OpenAI Codex (GPT-6 Astra) also assisted with supplementary
numerical cross-checks and the preparation of the figure-reproduction
scripts. These scripts were checked against the author's original
implementation for consistency of the pulse sequences and calculation
settings.

For grayscale maps we set $F_{\mathrm{cut}}=0.9999$ and use the
normalized intensity
\begin{equation}
  c(F)=\frac{\max(F,F_{\mathrm{cut}})-F_{\mathrm{cut}}}
              {1-F_{\mathrm{cut}}}.
  \label{eq:fidelity-color}
\end{equation}
The displayed gray level is $\lfloor255c(F)\rfloor$.
Black therefore denotes $F\leq F_{\mathrm{cut}}$, not zero fidelity,
and white denotes $F=1$. All compared panels use the same scale.

For a construction satisfying both first-order cancellation conditions,
the phase-aligned operation obeys
\begin{equation}
  W_0^\dagger W(\varepsilon,f)
  =I+O(\varepsilon^2)+O(\varepsilon f)+O(f^2).
  \label{eq:finite-remainder}
\end{equation}
The remaining higher-order and mixed-error terms depend on the
construction and determine its response at finite error strengths.

Figure~\ref{fig:fidelity-maps} illustrates one representative concatenation
from each inner-pulse family, together with reference sequences.
All six panels use the same target $R(\pi,0)$, error domain,
fidelity definition, and grayscale. For short CORPSE inner pulses,
CORPSE/BB1 provides a conventional local-REP reference, and
shortCORPSE/BB1 isolates the failure of direct replacement without
equal-angle splitting. For SCROFULOUS inner pulses, the two constituent
sequences provide controls that compensate one error type each.
The maps characterize finite-error performance; first-order robustness
is established separately by the analytical generators in
Sec.~\ref{sec:constructions}, not inferred from the appearance of a map.

We write concatenations as inner/outer: the
left-hand name specifies the composite blocks visible in the expanded
pulse sequence, while the right-hand name specifies the outer
sequence formed by those blocks.
The conventional CORPSE/BB1 construction in panel (a) has a
high-fidelity region extending along both error axes. Replacing
CORPSE by short CORPSE without splitting BB1 leaves a narrow band
in panel (b). Equal-angle splitting of BB1 restores the extent of
the high-fidelity region in the PLE direction in panel (c), consistent
with the cancellation established in Sec.~\ref{subsec:short-corpse}.

The lower row illustrates the complementary construction.
SCROFULOUS alone in panel (d) compensates PLE, whereas the split
short CORPSE outer sequence in panel (e) compensates ORE.
Splitting does not alter the error-dependent operation of short
CORPSE under Eq.~\eqref{eq:simultaneous-model}.
Their concatenation in panel (f) has a high-fidelity region with
finite extent in both directions near the origin, consistent with
the first-order cancellation in Sec.~\ref{subsec:scrofulous}.
Its contour shape differs from that of panel (c), reflecting the
different higher-order responses of the two constructions.
These illustrative comparisons do not establish a uniform performance
ordering, an advantage at equal implementation cost, or superiority
to direct simultaneous-error designs such as those in
Ref.~\cite{Jones2013}.

\subsection{Pulse count and total rotation angle}
\label{subsec:implementation-cost}

We report the costs of all four explicit concatenated constructions
in Sec.~\ref{sec:constructions}, including the split-SK1 and
split-CORPSE variants not plotted in Fig.~\ref{fig:fidelity-maps}.
Table~\ref{tab:construction-cost} summarizes these four constructions;
the reference sequences used in the figure are not included.

We count the elementary pulses after replacing each outer element by
its inner composite sequence. Let $M$ denote this count, and define
the total rotation angle by
\begin{equation}
  L=\sum_{\ell=1}^{M}\vartheta_\ell,
  \qquad T=\frac{L}{\Omega},
  \label{eq:implementation-cost}
\end{equation}
where all implemented angles $\vartheta_\ell$ are positive.
The duration $T$ is the error-free execution time at a common Rabi
angular frequency $\Omega$, with no intervening delays or switching
overhead. Neither $M$ nor $L$ alone determines fidelity.

A short CORPSE block implementing angle $\alpha$ contains three
pulses and has total angle
\begin{equation}
  L_{\mathrm{short}}(\alpha)
  =\alpha+2\pi-4\arcsin\left[\frac{\sin(\alpha/2)}{2}\right].
  \label{eq:short-cost}
\end{equation}
Thus an equal-angle outer sequence with $N$ elements gives
$M=3N$ and $L=N L_{\mathrm{short}}(\alpha)$ before merging.
If this outer sequence is obtained by splitting angles
$\theta_j=m_j\alpha$, then $N=\sum_j m_j$.
At $\alpha=\pi$, the short CORPSE angles are
$(\pi/3,5\pi/3,\pi/3)$ and $L_{\mathrm{short}}=7\pi/3$.
Both split-SK1 and split-BB1 examples in Sec.~\ref{subsec:short-corpse}
therefore have $M=15$ and $L=35\pi/3$.

On the SCROFULOUS branch of Sec.~\ref{subsec:scrofulous}, a block
has total angle $L_{\mathrm{SCROF}}(\alpha)=2a(\alpha)+\pi$,
where $a(\alpha)$ solves Eq.~\eqref{eq:scrof-angle}.
An $N$-element equal-angle outer sequence consequently gives
$M=3N$ and $L=N[2a(\alpha)+\pi]$ before merging.
For the split CORPSE-family examples, $\alpha=\pi/3$.
Write $a_*=a(\pi/3)$ and $\ell_*=1+2a_*/\pi$, where
$\sin a_*/a_*=\sqrt{3}/\pi$. The split short CORPSE outer
sequence has $N=7$, giving $M=21$ and $L=7\pi\ell_*$.
The split CORPSE outer sequence has $N=13$, giving $M=39$
and $L=13\pi\ell_*$.
Table~\ref{tab:construction-cost} summarizes these costs.

\begin{table}[t]
  \caption{Costs of the constructions in Sec.~\ref{sec:constructions}
  for a target $\pi$ rotation, before merging adjacent pulses.
  $L$ is the total rotation angle and $T=L/\Omega$;
  $\ell_*=1+2a(\pi/3)/\pi$.}
  \label{tab:construction-cost}
  \begin{ruledtabular}
  \begin{tabular}{llcc}
    Outer sequence & Inner sequence & $M$ & $L/\pi$ \\
    \hline
    Split SK1 & Short CORPSE & 15 & $35/3$ \\
    Split BB1 & Short CORPSE & 15 & $35/3$ \\
    Split short CORPSE & SCROFULOUS & 21 & $7\ell_*$ \\
    Split CORPSE & SCROFULOUS & 39 & $13\ell_*$
  \end{tabular}
  \end{ruledtabular}
\end{table}

Consecutive pulses of the same phase can be merged exactly under
Eq.~\eqref{eq:simultaneous-model}, provided their Rabi frequency and
systematic errors are shared and no delay intervenes:
\begin{equation}
  R(v,\psi;\varepsilon,f)R(u,\psi;\varepsilon,f)
  =R(u+v,\psi;\varepsilon,f).
  \label{eq:exact-merging}
\end{equation}
This reduces the pulse count without changing $L$ or the finite-error
operation. A phase difference of $\pi$ does not in general permit the
same simplification: it reverses the transverse drive but not the
detuning term. Costs before and after any merging must therefore be
distinguished. The values in Table~\ref{tab:construction-cost} describe
the explicit unmerged constructions and are not minimality claims.

\section{Discussion and Conclusions}
\label{sec:conclusions}

Local residual-error preservation is sufficient, but not necessary,
for concatenated composite pulses. Applying the established
first-order composition rule of Ref.~\cite{Ichikawa2011} to the
difference between the inner and elementary responses gives a
necessary and sufficient condition for retaining an outer sequence's
first-order robustness: deviations from local REP must cancel after
conjugation by the preceding ideal operations, rather than vanish
in every replacement. The criterion depends on the combined
residual responses, not on a particular control Hamiltonian.

A common real proportionality factor provides a simple sufficient
condition for this cancellation, without requiring preservation of
the magnitude or sign of each residual generator. Short CORPSE and
SCROFULOUS realize this structure with the roles of PLE and ORE
interchanged. Suitable equal-angle outer sequences yield simultaneous
first-order compensation in both cases; the short CORPSE example
with factor $q=-1$ explicitly demonstrates that even reversal of
the residual generator is compatible with cancellation.
Equal angles are a construction method for these families, not a
necessary condition for the general criterion.

The analysis concerns shared systematic errors and replacements
with matching ideal operations. The explicit constructions are
subject to the control-model assumptions, splitting conditions,
and inner-pulse domains specified in Sec.~\ref{sec:constructions}.
Higher-order and mixed-error terms, pulse-dependent errors, switching
transients, and decoherence are not covered by the first-order
guarantee. The fidelity maps and costs in Sec.~\ref{sec:performance}
illustrate finite-error responses and implementation requirements,
without establishing optimality. Choosing the splitting angle or
seeking unequal-angle outer sequences that satisfy the global
condition offers further scope to balance performance and cost.
The reduced-CCCP constructions of Ref.~\cite{Bando2013} avoid replacing
blocks that already compensate the error targeted by the inner pulses,
while local REP preserves cancellation of the other error.
Such selective replacement does not automatically preserve the
common-factor condition used here: unreplaced operations retain
their elementary response rather than acquiring the factor $q$.
The global criterion can test reductions beyond local REP, but finding
shorter implementations that retain simultaneous first-order
cancellation remains an open design problem.

The design principle is to match inner residual responses to outer
cancellation, rather than to elementary responses locally. This
enlarges the class of usable inner composite pulses and provides an
analytical test of compatibility with an outer sequence.

\begin{acknowledgments}
This work was supported by internal research funding from
Kindai University Technical College.

The author used OpenAI Codex (GPT-6 Astra) to assist with
English-language drafting and revision and with refining the
presentation of analytical results.
The research ideas, analytical derivations, and overall
manuscript structure were developed by the author. The author directed
the AI assistance, checked the mathematical content against his own
derivations, and reviewed and revised the text, taking full responsibility
for the final manuscript.
\end{acknowledgments}

\section*{Data Availability}
The numerical library is publicly available on GitHub as
\texttt{v0.1.13}~\cite{BandoQuantumGates2026} and on npm as
\path{@masabando/quantum-gates@0.1.13}.
Code reproducing all six panels of Fig.~\ref{fig:fidelity-maps}, with
pulse definitions, settings, and instructions, is available in
\texttt{cccp-beyond-rep}, tag \texttt{v1.0.1}~\cite{BandoReproduction2026},
using this library version. A browser-based interface is available at
\url{https://masabando.github.io/cccp-beyond-rep/}.
The fidelity values are generated by this code;
no external input dataset is required.

\bibliography{references}

\begin{thebibliography}{23}%
\makeatletter
\providecommand \@ifxundefined [1]{%
 \@ifx{#1\undefined}
}%
\providecommand \@ifnum [1]{%
 \ifnum #1\expandafter \@firstoftwo
 \else \expandafter \@secondoftwo
 \fi
}%
\providecommand \@ifx [1]{%
 \ifx #1\expandafter \@firstoftwo
 \else \expandafter \@secondoftwo
 \fi
}%
\providecommand \natexlab [1]{#1}%
\providecommand \enquote  [1]{``#1''}%
\providecommand \bibnamefont  [1]{#1}%
\providecommand \bibfnamefont [1]{#1}%
\providecommand \citenamefont [1]{#1}%
\providecommand \href@noop [0]{\@secondoftwo}%
\providecommand \href [0]{\begingroup \@sanitize@url \@href}%
\providecommand \@href[1]{\@@startlink{#1}\@@href}%
\providecommand \@@href[1]{\endgroup#1\@@endlink}%
\providecommand \@sanitize@url [0]{\catcode `\\12\catcode `\$12\catcode
  `\&12\catcode `\#12\catcode `\^12\catcode `\_12\catcode `\%12\relax}%
\providecommand \@@startlink[1]{}%
\providecommand \@@endlink[0]{}%
\providecommand \url  [0]{\begingroup\@sanitize@url \@url }%
\providecommand \@url [1]{\endgroup\@href {#1}{\urlprefix }}%
\providecommand \urlprefix  [0]{URL }%
\providecommand \Eprint [0]{\href }%
\providecommand \doibase [0]{https://doi.org/}%
\providecommand \selectlanguage [0]{\@gobble}%
\providecommand \bibinfo  [0]{\@secondoftwo}%
\providecommand \bibfield  [0]{\@secondoftwo}%
\providecommand \translation [1]{[#1]}%
\providecommand \BibitemOpen [0]{}%
\providecommand \bibitemStop [0]{}%
\providecommand \bibitemNoStop [0]{.\EOS\space}%
\providecommand \EOS [0]{\spacefactor3000\relax}%
\providecommand \BibitemShut  [1]{\csname bibitem#1\endcsname}%
\let\auto@bib@innerbib\@empty
%</preamble>
\bibitem [{\citenamefont {Levitt}(1986)}]{Levitt1986}%
  \BibitemOpen
  \bibfield  {author} {\bibinfo {author} {\bibfnamefont {M.~H.}\ \bibnamefont
  {Levitt}},\ }\bibfield  {title} {\bibinfo {title} {Composite pulses},\ }\href
  {https://doi.org/10.1016/0079-6565(86)80005-X} {\bibfield  {journal}
  {\bibinfo  {journal} {Prog. Nucl. Magn. Reson. Spectrosc.}\ }\textbf
  {\bibinfo {volume} {18}},\ \bibinfo {pages} {61} (\bibinfo {year}
  {1986})}\BibitemShut {NoStop}%
\bibitem [{\citenamefont {Cummins}\ \emph {et~al.}(2003)\citenamefont
  {Cummins}, \citenamefont {Llewellyn},\ and\ \citenamefont
  {Jones}}]{Cummins2003}%
  \BibitemOpen
  \bibfield  {author} {\bibinfo {author} {\bibfnamefont {H.~K.}\ \bibnamefont
  {Cummins}}, \bibinfo {author} {\bibfnamefont {G.}~\bibnamefont {Llewellyn}},\
  and\ \bibinfo {author} {\bibfnamefont {J.~A.}\ \bibnamefont {Jones}},\
  }\bibfield  {title} {\bibinfo {title} {Tackling systematic errors in quantum
  logic gates with composite rotations},\ }\href
  {https://doi.org/10.1103/PhysRevA.67.042308} {\bibfield  {journal} {\bibinfo
  {journal} {Phys. Rev. A}\ }\textbf {\bibinfo {volume} {67}},\ \bibinfo
  {pages} {042308} (\bibinfo {year} {2003})}\BibitemShut {NoStop}%
\bibitem [{\citenamefont {Cummins}\ and\ \citenamefont
  {Jones}(2000)}]{Cummins2000}%
  \BibitemOpen
  \bibfield  {author} {\bibinfo {author} {\bibfnamefont {H.~K.}\ \bibnamefont
  {Cummins}}\ and\ \bibinfo {author} {\bibfnamefont {J.~A.}\ \bibnamefont
  {Jones}},\ }\bibfield  {title} {\bibinfo {title} {Use of composite rotations
  to correct systematic errors in {NMR} quantum computation},\ }\href
  {https://doi.org/10.1088/1367-2630/2/1/006} {\bibfield  {journal} {\bibinfo
  {journal} {New J. Phys.}\ }\textbf {\bibinfo {volume} {2}},\ \bibinfo {pages}
  {6} (\bibinfo {year} {2000})}\BibitemShut {NoStop}%
\bibitem [{\citenamefont {Jones}(2011)}]{Jones2011}%
  \BibitemOpen
  \bibfield  {author} {\bibinfo {author} {\bibfnamefont {J.~A.}\ \bibnamefont
  {Jones}},\ }\bibfield  {title} {\bibinfo {title} {Quantum computing with
  {NMR}},\ }\href {https://doi.org/10.1016/j.pnmrs.2010.11.001} {\bibfield
  {journal} {\bibinfo  {journal} {Prog. Nucl. Magn. Reson. Spectrosc.}\
  }\textbf {\bibinfo {volume} {59}},\ \bibinfo {pages} {91} (\bibinfo {year}
  {2011})}\BibitemShut {NoStop}%
\bibitem [{\citenamefont {Wimperis}(1994)}]{Wimperis1994}%
  \BibitemOpen
  \bibfield  {author} {\bibinfo {author} {\bibfnamefont {S.}~\bibnamefont
  {Wimperis}},\ }\bibfield  {title} {\bibinfo {title} {Broadband, narrowband,
  and passband composite pulses for use in advanced {NMR} experiments},\ }\href
  {https://doi.org/10.1006/jmra.1994.1159} {\bibfield  {journal} {\bibinfo
  {journal} {J. Magn. Reson. A}\ }\textbf {\bibinfo {volume} {109}},\ \bibinfo
  {pages} {221} (\bibinfo {year} {1994})}\BibitemShut {NoStop}%
\bibitem [{\citenamefont {Brown}\ \emph {et~al.}(2004)\citenamefont {Brown},
  \citenamefont {Harrow},\ and\ \citenamefont {Chuang}}]{Brown2004}%
  \BibitemOpen
  \bibfield  {author} {\bibinfo {author} {\bibfnamefont {K.~R.}\ \bibnamefont
  {Brown}}, \bibinfo {author} {\bibfnamefont {A.~W.}\ \bibnamefont {Harrow}},\
  and\ \bibinfo {author} {\bibfnamefont {I.~L.}\ \bibnamefont {Chuang}},\
  }\bibfield  {title} {\bibinfo {title} {Arbitrarily accurate composite pulse
  sequences},\ }\href {https://doi.org/10.1103/PhysRevA.70.052318} {\bibfield
  {journal} {\bibinfo  {journal} {Phys. Rev. A}\ }\textbf {\bibinfo {volume}
  {70}},\ \bibinfo {pages} {052318} (\bibinfo {year} {2004})},\ \bibinfo {note}
  {erratum: Phys. Rev. A 72, 039905 (2005)}\BibitemShut {NoStop}%
\bibitem [{\citenamefont {Alway}\ and\ \citenamefont
  {Jones}(2007)}]{Alway2007}%
  \BibitemOpen
  \bibfield  {author} {\bibinfo {author} {\bibfnamefont {W.~G.}\ \bibnamefont
  {Alway}}\ and\ \bibinfo {author} {\bibfnamefont {J.~A.}\ \bibnamefont
  {Jones}},\ }\bibfield  {title} {\bibinfo {title} {Arbitrary precision
  composite pulses for {NMR} quantum computing},\ }\href
  {https://doi.org/10.1016/j.jmr.2007.09.001} {\bibfield  {journal} {\bibinfo
  {journal} {J. Magn. Reson.}\ }\textbf {\bibinfo {volume} {189}},\ \bibinfo
  {pages} {114} (\bibinfo {year} {2007})}\BibitemShut {NoStop}%
\bibitem [{\citenamefont {Low}\ \emph {et~al.}(2014)\citenamefont {Low},
  \citenamefont {Yoder},\ and\ \citenamefont {Chuang}}]{Low2014}%
  \BibitemOpen
  \bibfield  {author} {\bibinfo {author} {\bibfnamefont {G.~H.}\ \bibnamefont
  {Low}}, \bibinfo {author} {\bibfnamefont {T.~J.}\ \bibnamefont {Yoder}},\
  and\ \bibinfo {author} {\bibfnamefont {I.~L.}\ \bibnamefont {Chuang}},\
  }\bibfield  {title} {\bibinfo {title} {Optimal arbitrarily accurate composite
  pulse sequences},\ }\href {https://doi.org/10.1103/PhysRevA.89.022341}
  {\bibfield  {journal} {\bibinfo  {journal} {Phys. Rev. A}\ }\textbf {\bibinfo
  {volume} {89}},\ \bibinfo {pages} {022341} (\bibinfo {year}
  {2014})}\BibitemShut {NoStop}%
\bibitem [{\citenamefont {Ichikawa}\ \emph {et~al.}(2011)\citenamefont
  {Ichikawa}, \citenamefont {Bando}, \citenamefont {Kondo},\ and\ \citenamefont
  {Nakahara}}]{Ichikawa2011}%
  \BibitemOpen
  \bibfield  {author} {\bibinfo {author} {\bibfnamefont {T.}~\bibnamefont
  {Ichikawa}}, \bibinfo {author} {\bibfnamefont {M.}~\bibnamefont {Bando}},
  \bibinfo {author} {\bibfnamefont {Y.}~\bibnamefont {Kondo}},\ and\ \bibinfo
  {author} {\bibfnamefont {M.}~\bibnamefont {Nakahara}},\ }\bibfield  {title}
  {\bibinfo {title} {Designing robust unitary gates: Application to
  concatenated composite pulses},\ }\href
  {https://doi.org/10.1103/PhysRevA.84.062311} {\bibfield  {journal} {\bibinfo
  {journal} {Phys. Rev. A}\ }\textbf {\bibinfo {volume} {84}},\ \bibinfo
  {pages} {062311} (\bibinfo {year} {2011})}\BibitemShut {NoStop}%
\bibitem [{\citenamefont {Bando}\ \emph {et~al.}(2013)\citenamefont {Bando},
  \citenamefont {Ichikawa}, \citenamefont {Kondo},\ and\ \citenamefont
  {Nakahara}}]{Bando2013}%
  \BibitemOpen
  \bibfield  {author} {\bibinfo {author} {\bibfnamefont {M.}~\bibnamefont
  {Bando}}, \bibinfo {author} {\bibfnamefont {T.}~\bibnamefont {Ichikawa}},
  \bibinfo {author} {\bibfnamefont {Y.}~\bibnamefont {Kondo}},\ and\ \bibinfo
  {author} {\bibfnamefont {M.}~\bibnamefont {Nakahara}},\ }\bibfield  {title}
  {\bibinfo {title} {Concatenated composite pulses compensating simultaneous
  systematic errors},\ }\href {https://doi.org/10.7566/JPSJ.82.014004}
  {\bibfield  {journal} {\bibinfo  {journal} {J. Phys. Soc. Jpn.}\ }\textbf
  {\bibinfo {volume} {82}},\ \bibinfo {pages} {014004} (\bibinfo {year}
  {2013})}\BibitemShut {NoStop}%
\bibitem [{\citenamefont {Bando}\ \emph {et~al.}(2020)\citenamefont {Bando},
  \citenamefont {Ichikawa}, \citenamefont {Kondo}, \citenamefont {Nemoto},
  \citenamefont {Nakahara},\ and\ \citenamefont {Shikano}}]{Bando2020}%
  \BibitemOpen
  \bibfield  {author} {\bibinfo {author} {\bibfnamefont {M.}~\bibnamefont
  {Bando}}, \bibinfo {author} {\bibfnamefont {T.}~\bibnamefont {Ichikawa}},
  \bibinfo {author} {\bibfnamefont {Y.}~\bibnamefont {Kondo}}, \bibinfo
  {author} {\bibfnamefont {N.}~\bibnamefont {Nemoto}}, \bibinfo {author}
  {\bibfnamefont {M.}~\bibnamefont {Nakahara}},\ and\ \bibinfo {author}
  {\bibfnamefont {Y.}~\bibnamefont {Shikano}},\ }\bibfield  {title} {\bibinfo
  {title} {Concatenated composite pulses applied to liquid-state nuclear
  magnetic resonance spectroscopy},\ }\href
  {https://doi.org/10.1038/s41598-020-58823-9} {\bibfield  {journal} {\bibinfo
  {journal} {Sci. Rep.}\ }\textbf {\bibinfo {volume} {10}},\ \bibinfo {pages}
  {2126} (\bibinfo {year} {2020})}\BibitemShut {NoStop}%
\bibitem [{\citenamefont {Odedra}\ \emph {et~al.}(2012)\citenamefont {Odedra},
  \citenamefont {Thrippleton},\ and\ \citenamefont {Wimperis}}]{Odedra2012}%
  \BibitemOpen
  \bibfield  {author} {\bibinfo {author} {\bibfnamefont {S.}~\bibnamefont
  {Odedra}}, \bibinfo {author} {\bibfnamefont {M.~J.}\ \bibnamefont
  {Thrippleton}},\ and\ \bibinfo {author} {\bibfnamefont {S.}~\bibnamefont
  {Wimperis}},\ }\bibfield  {title} {\bibinfo {title} {Dual-compensated
  antisymmetric composite refocusing pulses for {NMR}},\ }\href
  {https://doi.org/10.1016/j.jmr.2012.10.003} {\bibfield  {journal} {\bibinfo
  {journal} {J. Magn. Reson.}\ }\textbf {\bibinfo {volume} {225}},\ \bibinfo
  {pages} {81} (\bibinfo {year} {2012})}\BibitemShut {NoStop}%
\bibitem [{\citenamefont {Jones}(2013{\natexlab{a}})}]{Jones2013}%
  \BibitemOpen
  \bibfield  {author} {\bibinfo {author} {\bibfnamefont {J.~A.}\ \bibnamefont
  {Jones}},\ }\bibfield  {title} {\bibinfo {title} {Designing short robust
  {NOT} gates for quantum computation},\ }\href
  {https://doi.org/10.1103/PhysRevA.87.052317} {\bibfield  {journal} {\bibinfo
  {journal} {Phys. Rev. A}\ }\textbf {\bibinfo {volume} {87}},\ \bibinfo
  {pages} {052317} (\bibinfo {year} {2013}{\natexlab{a}})}\BibitemShut
  {NoStop}%
\bibitem [{\citenamefont {Ichikawa}\ \emph {et~al.}(2014)\citenamefont
  {Ichikawa}, \citenamefont {Filgueiras}, \citenamefont {Bando}, \citenamefont
  {Kondo}, \citenamefont {Nakahara},\ and\ \citenamefont
  {Suter}}]{Ichikawa2014}%
  \BibitemOpen
  \bibfield  {author} {\bibinfo {author} {\bibfnamefont {T.}~\bibnamefont
  {Ichikawa}}, \bibinfo {author} {\bibfnamefont {J.~G.}\ \bibnamefont
  {Filgueiras}}, \bibinfo {author} {\bibfnamefont {M.}~\bibnamefont {Bando}},
  \bibinfo {author} {\bibfnamefont {Y.}~\bibnamefont {Kondo}}, \bibinfo
  {author} {\bibfnamefont {M.}~\bibnamefont {Nakahara}},\ and\ \bibinfo
  {author} {\bibfnamefont {D.}~\bibnamefont {Suter}},\ }\bibfield  {title}
  {\bibinfo {title} {Construction of arbitrary robust one-qubit operations
  using planar geometry},\ }\href {https://doi.org/10.1103/PhysRevA.90.052330}
  {\bibfield  {journal} {\bibinfo  {journal} {Phys. Rev. A}\ }\textbf {\bibinfo
  {volume} {90}},\ \bibinfo {pages} {052330} (\bibinfo {year}
  {2014})}\BibitemShut {NoStop}%
\bibitem [{\citenamefont {Tonchev}\ and\ \citenamefont
  {Vitanov}(2026)}]{Tonchev2026}%
  \BibitemOpen
  \bibfield  {author} {\bibinfo {author} {\bibfnamefont {H.~G.}\ \bibnamefont
  {Tonchev}}\ and\ \bibinfo {author} {\bibfnamefont {N.~V.}\ \bibnamefont
  {Vitanov}},\ }\href@noop {} {\bibinfo {title} {Composite quantum gates
  simultaneously compensated for multiple errors}} (\bibinfo {year} {2026}),\
  \Eprint {https://arxiv.org/abs/2604.21594} {arXiv:2604.21594 [quant-ph]}
  \BibitemShut {NoStop}%
\bibitem [{\citenamefont {Jones}(2013{\natexlab{b}})}]{JonesNested2013}%
  \BibitemOpen
  \bibfield  {author} {\bibinfo {author} {\bibfnamefont {J.~A.}\ \bibnamefont
  {Jones}},\ }\bibfield  {title} {\bibinfo {title} {Nested composite {NOT}
  gates for quantum computation},\ }\href
  {https://doi.org/10.1016/j.physleta.2013.08.040} {\bibfield  {journal}
  {\bibinfo  {journal} {Phys. Lett. A}\ }\textbf {\bibinfo {volume} {377}},\
  \bibinfo {pages} {2860} (\bibinfo {year} {2013}{\natexlab{b}})}\BibitemShut
  {NoStop}%
\bibitem [{\citenamefont {M{\"o}tt{\"o}nen}\ \emph {et~al.}(2006)\citenamefont
  {M{\"o}tt{\"o}nen}, \citenamefont {de~Sousa}, \citenamefont {Zhang},\ and\
  \citenamefont {Whaley}}]{Mottonen2006}%
  \BibitemOpen
  \bibfield  {author} {\bibinfo {author} {\bibfnamefont {M.}~\bibnamefont
  {M{\"o}tt{\"o}nen}}, \bibinfo {author} {\bibfnamefont {R.}~\bibnamefont
  {de~Sousa}}, \bibinfo {author} {\bibfnamefont {J.}~\bibnamefont {Zhang}},\
  and\ \bibinfo {author} {\bibfnamefont {K.~B.}\ \bibnamefont {Whaley}},\
  }\bibfield  {title} {\bibinfo {title} {High-fidelity one-qubit operations
  under random telegraph noise},\ }\href
  {https://doi.org/10.1103/PhysRevA.73.022332} {\bibfield  {journal} {\bibinfo
  {journal} {Phys. Rev. A}\ }\textbf {\bibinfo {volume} {73}},\ \bibinfo
  {pages} {022332} (\bibinfo {year} {2006})}\BibitemShut {NoStop}%
\bibitem [{\citenamefont {Khodjasteh}\ and\ \citenamefont
  {Viola}(2009)}]{Khodjasteh2009}%
  \BibitemOpen
  \bibfield  {author} {\bibinfo {author} {\bibfnamefont {K.}~\bibnamefont
  {Khodjasteh}}\ and\ \bibinfo {author} {\bibfnamefont {L.}~\bibnamefont
  {Viola}},\ }\bibfield  {title} {\bibinfo {title} {Dynamical quantum error
  correction of unitary operations with bounded controls},\ }\href
  {https://doi.org/10.1103/PhysRevA.80.032314} {\bibfield  {journal} {\bibinfo
  {journal} {Phys. Rev. A}\ }\textbf {\bibinfo {volume} {80}},\ \bibinfo
  {pages} {032314} (\bibinfo {year} {2009})}\BibitemShut {NoStop}%
\bibitem [{\citenamefont {Ryan}\ \emph {et~al.}(2010)\citenamefont {Ryan},
  \citenamefont {Hodges},\ and\ \citenamefont {Cory}}]{Ryan2010}%
  \BibitemOpen
  \bibfield  {author} {\bibinfo {author} {\bibfnamefont {C.~A.}\ \bibnamefont
  {Ryan}}, \bibinfo {author} {\bibfnamefont {J.~S.}\ \bibnamefont {Hodges}},\
  and\ \bibinfo {author} {\bibfnamefont {D.~G.}\ \bibnamefont {Cory}},\
  }\bibfield  {title} {\bibinfo {title} {Robust decoupling techniques to extend
  quantum coherence in diamond},\ }\href
  {https://doi.org/10.1103/PhysRevLett.105.200402} {\bibfield  {journal}
  {\bibinfo  {journal} {Phys. Rev. Lett.}\ }\textbf {\bibinfo {volume} {105}},\
  \bibinfo {pages} {200402} (\bibinfo {year} {2010})}\BibitemShut {NoStop}%
\bibitem [{\citenamefont {Souza}\ \emph {et~al.}(2012)\citenamefont {Souza},
  \citenamefont {{\'{A}lvarez}},\ and\ \citenamefont {Suter}}]{Souza2012}%
  \BibitemOpen
  \bibfield  {author} {\bibinfo {author} {\bibfnamefont {A.~M.}\ \bibnamefont
  {Souza}}, \bibinfo {author} {\bibfnamefont {G.~A.}\ \bibnamefont
  {{\'{A}lvarez}}},\ and\ \bibinfo {author} {\bibfnamefont {D.}~\bibnamefont
  {Suter}},\ }\bibfield  {title} {\bibinfo {title} {Experimental protection of
  quantum gates against decoherence and control errors},\ }\href
  {https://doi.org/10.1103/PhysRevA.86.050301} {\bibfield  {journal} {\bibinfo
  {journal} {Phys. Rev. A}\ }\textbf {\bibinfo {volume} {86}},\ \bibinfo
  {pages} {050301(R)} (\bibinfo {year} {2012})}\BibitemShut {NoStop}%
\bibitem [{\citenamefont {Claridge}(1999)}]{Claridge1999}%
  \BibitemOpen
  \bibfield  {author} {\bibinfo {author} {\bibfnamefont {T.~D.~W.}\
  \bibnamefont {Claridge}},\ }\href@noop {} {\emph {\bibinfo {title}
  {High-Resolution {NMR} Techniques in Organic Chemistry}}}\ (\bibinfo
  {publisher} {Elsevier},\ \bibinfo {address} {Amsterdam},\ \bibinfo {year}
  {1999})\BibitemShut {NoStop}%
\bibitem [{\citenamefont {Bando}(2026{\natexlab{a}})}]{BandoQuantumGates2026}%
  \BibitemOpen
  \bibfield  {author} {\bibinfo {author} {\bibfnamefont {M.}~\bibnamefont
  {Bando}},\ }\href {https://github.com/masabando/quantum-gates/tree/v0.1.13}
  {\bibinfo {title} {{@masabando/quantum-gates}}},\ \bibinfo {howpublished}
  {GitHub} (\bibinfo {year} {2026}{\natexlab{a}}),\ \bibinfo {note} {version
  0.1.13 [Software],
  \url{https://github.com/masabando/quantum-gates/tree/v0.1.13}}\BibitemShut
  {NoStop}%
\bibitem [{\citenamefont {Bando}(2026{\natexlab{b}})}]{BandoReproduction2026}%
  \BibitemOpen
  \bibfield  {author} {\bibinfo {author} {\bibfnamefont {M.}~\bibnamefont
  {Bando}},\ }\href {https://github.com/masabando/cccp-beyond-rep/tree/v1.0.1}
  {\bibinfo {title} {{cccp-beyond-rep}: Figure reproduction code for
  concatenated composite pulses beyond local residual-error preservation}},\
  \bibinfo {howpublished} {GitHub} (\bibinfo {year} {2026}{\natexlab{b}}),\
  \bibinfo {note} {version 1.0.1 [Software],
  \url{https://github.com/masabando/cccp-beyond-rep/tree/v1.0.1}}\BibitemShut
  {NoStop}%
\end{thebibliography}%

\end{document}